\documentclass{elsart3}
\usepackage{amssymb, amsmath}
\usepackage[dvips]{graphicx}

\begin{document}

\begin{frontmatter}

\title{Intensity of Coulomb interaction between quasiparticles \\ in diffusive metallic wires}
\author[Saclay]{B. Huard},
\author[Saclay]{A. Anthore\thanksref{Helsinki}},
\author[Saclay,MSU]{F. Pierre\thanksref{LPN}},
\author[Saclay]{H. Pothier\corauthref{Hugues}}
\ead{hpothier@cea.fr},
\author[Saclay,MSU]{Norman O. Birge},
\author[Saclay]{D. Esteve}
\address[Saclay]{Quantronics Group, Service de Physique de l'Etat Condens\'{e}, DRECAM, CEA-Saclay, 91191 Gif-sur-Yvette, France}
\address[MSU]{Department of Physics and Astronomy, Michigan State University, East Lansing, MI 48824, USA}
\thanks[Helsinki]{Present address : Low Temperature Laboratory, Helsinki University of Technology, PO Box 3500, FIN - 02015, HUT, Finland}
\thanks[LPN]{Present address : Laboratoire de Photonique et Nanostructures, CNRS, Route de Nozay, 91460 Marcoussis, France}
\corauth[Hugues]{Corresponding author}

\begin{abstract}
The energy dependence and intensity of Coulomb interaction between quasiparticles in metallic wires is obtained from two different methods~: determination of the temperature dependence of the phase coherence time from the magnetoresistance, and measurements of the energy distribution function in out-of-equilibrium situations. In both types of experiment, the energy dependence of the Coulomb interaction is found to be in excellent agreement with theoretical predictions.  In contrast, the intensity of the interaction agrees closely with theory only with the first method, whereas an important discrepancy is found using the second one. Different explanations are proposed, and results of a test experiment are presented.
\end{abstract}

\end{frontmatter}

\section{Introduction}

The description of electrical transport in metals is based on the
existence of long-lived quasiparticles. The finite quasiparticle
lifetime appears in mesoscopic physics as a limitation of their
phase coherence time, which determines the amplitude of quantum
interference effects.  The three kinds of processes that limit the
quasiparticle lifetime in metals are electron-phonon scattering,
electron-electron scattering \cite{AA}, and spin-flip scattering of
electrons from magnetic impurities \cite{Hikami,KG}.  At
temperatures below about 1~K, the rate of electron-phonon
scattering is weak, and in metallic samples without magnetic
impurities the dominant inelastic scattering process should be the
Coulomb interaction between electrons \cite{AA}.

In this paper, we focus on experiments performed on very clean
(99.9999\%) silver wires, in which the effect of magnetic
impurities is expected to be small \cite{PRB,PRLAnne}. We review the results obtained from weak
localization measurements, in which the phase coherence time
$\tau_{\varphi}(T)$ is extracted, and from energy relaxation
experiments, in which the energy exchange rate between
quasiparticles is derived from their energy distribution function
$f(E)$.  In the former experiments, we find that both the
temperature dependence and overall magnitude of
$\tau_{\varphi}(T)$ agree with the theoretical predictions.  In
the latter experiments, the energy dependence of the inelastic
rate agrees with theoretical predictions, but the overall
magnitude fluctuates significantly from sample to sample.

\section{Two experiments for measuring Coulomb interaction between QPs}

In metallic thin films, quasiparticles (QPs) experience frequent
elastic scattering from grain boundaries, film edges and
impurities. In this diffusive regime, characterized by a diffusion
constant $D$, the screening of the Coulomb interaction is retarded, and the
corresponding (squared) matrix element between two QPs, derived by
Altshuler and Aronov in the early 80s \cite{AA}, depends on the
energy $\varepsilon$ exchanged during the interaction process:
$\left| M(\varepsilon)\right| ^{2}\propto\varepsilon^{-3/2}$ in quasi-one-dimensional wires. This
energy dependence results in a temperature dependence of the phase
coherence time $\tau_{\varphi}(T)\propto T^{-2/3}$ \cite{AAK},
which has been observed in aluminum and silver wires by Wind \textit{et al.
}\cite{Wind} down to 1K, and by Echternach \textit{et al.\
}\cite{Echternach} in gold wires down to 100mK. The most convenient
method to access $\tau_{\varphi}$ is the measurement of the
magnetoresistance of wires with a length $L$ long compared to the
phase coherence length $L_{\varphi}=\sqrt{D\tau_{\varphi}},$ which
exhibits a small peak or dip at zero magnetic field due to weak
localization \cite{WLreview}. When the rate of spin precession
due to spin-orbit coupling exceeds the dephasing rate, as is
usually the case at low temperature, the relative amplitude of the
zero-field dip in the resistance gives direct access to
$L_{\varphi}:$
\begin{equation*}
\frac{\delta R}{R}\approx -\frac{2R}{R_{K}}\frac{L_{\varphi}}{L}
\end{equation*}
with $R_{K}=h/e^{2}\approx26~$k$\Omega$ the resistance quantum. The width in field of this dip corresponds to a flux quantum in the area $L_{\varphi}w,$ with $w$ the wire width. In practice, magnetoresistance curves measured at different temperatures are fit with a theoretical expression for $\frac{\delta R}{R}(B)$ in which the only fit parameters are the phase coherence length $L_{\varphi},$ the spin-orbit length $L_{so},$ and the width of the wire $w$ \cite{PRB}. The two last parameters, $L_{so}$ and $w,$ are fixed at a constant value independent of temperature for each sample \cite{Lso}.  Then, $\tau_{\varphi}$ is obtained as $L_{\varphi}^{2}/D,$ with $D$ obtained from the resistance $R=\frac{1}{\nu_{F}e^{2}D}\frac{L}{wt}$ where $\nu_{F}$ is the density of states at the Fermi energy (2 spin directions) and $t$ the wire thickness. In order to compare with theory, the resulting curve $\tau_{\varphi}(T)$ is fit with
\begin{equation}
\tau_{\varphi}(T)=(AT^{2/3}+BT^{3})^{-1}.   \label{tauphi}
\end{equation}
where $AT^{2/3}$ is the Coulomb interaction rate and $BT^{3}$ the approximate electron-phonon scattering rate \cite{SergeevMitin}.

In theory, the exchange part of the Coulomb interaction leads to \cite{AleinerWav}
\begin{equation}
A=\frac{1}{\hbar}\left( \frac{\pi k_{B}^{2}}{4\nu_{F}Lwt}\frac{R}{R_{K}}%
\right) ^{1/3}.   \label{A}
\end{equation}
The contribution due to the Hartree term has not been evaluated for wires \cite{Aleinerfilms}.

\begin{figure}[ptbh]
\begin{center}
\includegraphics[width=3.0in]{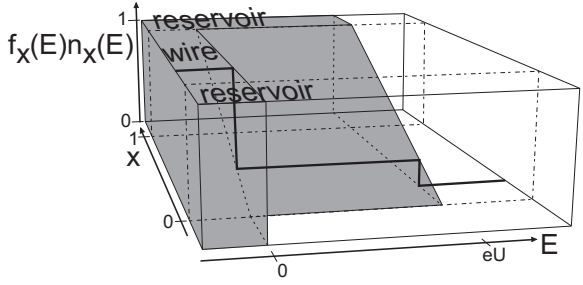} \caption{Schematic diagram showing the spatial and energy dependence of the distribution function $f_x(E)$ of QPs driven out-of-equilibrium by the voltage $U$ using the geometry of \textsc{Fig.} \ref{INJ scheme} with the switch in position 1. The surrounding box shows the uniform density of states in the metal and the gray volume shows the occupied states whose normalized density is $f_x(E)n_x(E)$. The thick line shows a typical double step distribution function at $x=1/4$ as in \textsc{Fig.}~\ref{INJ cas standard}.}\label{nf cas standard}
\end{center}
\end{figure}

Another experimental method to access the interaction processes
consists in driving the QPs out-of-equilibrium by a finite voltage
$U$ between two contacts at the ends of the wire, which act as QP reservoirs \cite{PRLrelax}.
At energies between $-eU$ and 0, the diffusion of QPs from the
occupied states at one end to empty states at the other end
results, in absence of inelastic processes, in a two-step
distribution function $f_{x}(E)$ inside the wire as pictured in
\textsc{Fig.} \ref{nf cas standard}.  (The shorthand $f_{x}(E)$
stands for $f(x,E)$, where we measure distance in units of the
wire length $L$, so that $0 < x < 1$.)  This distribution function
can be understood as a linear interpolation between the
distribution functions at the boundaries of the wire.
Electron-electron interactions lead to a redistribution of energy between QPs at each position, hence to a rounding of $f_{x}(E).$ In experiments, $f_{x}(E)$ at a given position in the wire is deduced from the differential conductance $dI/dV(V)$ of a tunnel junction between a superconducting probe electrode and the wire. In order to relate $f_{x}(E)$ to the matrix element of the interaction, the data are fit with the solution of the stationary Boltzmann equation in the diffusive regime \cite{Nagaev,Kozub}:
\begin{equation}
\frac{1}{\tau_{D}}\frac{\partial^{2}f_{x}\left(E\right)}{\partial x^{2}}+\mathcal{I}_{\mathrm{coll}}^{\mathrm{in}}\left( x,E,\left\{ f\right\}\right) -\mathcal{I}_{\mathrm{coll}}^{\mathrm{out}}\left( x,E,\left\{f\right\} \right)=0 \label{Boltzmann}
\end{equation}
where $\mathcal{I}_{\mathrm{coll}}^{\mathrm{in}}\left( x,E,\left\{ f\right\}
\right) $ and $\mathcal{I}_{\mathrm{coll}}^{\mathrm{out}}\left( x,E,\left\{
f\right\} \right) $ are the rates at which quasiparticles are scattered in and out of a state at energy $E$ by inelastic processes. The diffusion time $\tau_{D}=L^{2}/D$ is the typical time spent by a QP in the wire. Assuming that the dominant inelastic process is Coulomb interaction between QPs and phonon emission or absorption, the inelastic scattering integrals read
\begin{align*}
\mathcal{I}_{\mathrm{coll}}^{\mathrm{out}}\left( x,E,\left\{ f\right\}\right) & =\int\d\varepsilon~f_{x}(E)\left( 1-f_{x}(E-\varepsilon)\right) W(\varepsilon) \\\mathcal{I}_{\mathrm{coll}}^{\mathrm{in}}\left( x,E,\left\{ f\right\} \right) & =\int\d\varepsilon~f_{x}(E+\varepsilon)\left( 1-f_{x}(E)\right) W(\varepsilon)
\end{align*}
with
\begin{align*}
W(\varepsilon) & =W_{e-e}(\varepsilon)+W_{e-ph}(\varepsilon) \\
W_{e-e}(\varepsilon) & =K\left( \varepsilon\right) \int\d E^{\prime} f_{x}(E^{\prime})(1-f_{x}(E^{\prime}+\varepsilon))\\
W_{e-ph}(\varepsilon)&=\kappa_{ph}\varepsilon^{2}(n_{ph}(\left| \varepsilon\right| )+\theta(\varepsilon)).
\end{align*}

The kernel function $K\left( \varepsilon\right) =$
$\kappa_{ee}\varepsilon^{-3/2}$ is proportional to the
averaged squared interaction matrix element $\left|
M(\varepsilon)\right| ^{2}$ between two quasiparticles exchanging
an energy $\varepsilon$ \cite{AA}. Its intensity $\kappa_{ee}$,
which can be derived either from the expression of the microscopic
interaction potential \cite{Gilles,These Anne}, or from the
fluctuation-dissipation theorem \cite{These Anne}, is
\cite{factor2}
\begin{equation}
\kappa_{ee}=\left( \sqrt{2D}\pi\hbar^{3/2}\nu_{F}wt\right) ^{-1}.
\label{kappa}
\end{equation}
This derivation takes into account the exchange term only. The Hartree contribution to $K\left( \varepsilon\right)$ is expected to be smaller \cite{AA,Gilles}. The electron-phonon
coupling has an intensity $\kappa_{ph}$ and is proportional to the
sum of the Bose energy distribution of phonons $n_{ph}(\left|
\varepsilon\right| )$ representing stimulated absorption or
emission of phonons and the Heaviside function
$\theta(\varepsilon)$ representing spontaneous emission. A more
accurate description of electron-phonon coupling was developed in
\cite{SergeevMitin}. However, we restrict here to the simplistic
form for $W_{e-ph}$ because the effect of phonons is very
small. Thus, for all the fits to the experiments, we fix the value of $\kappa_{ph}$ at $4~$ns$^{-1}$meV$^{-3}$, which is compatible with the weak localization measurements\cite{kappaph facteur 2}.

The boundary conditions for \textsc{Eq.}~(\ref{Boltzmann}) are Fermi-Dirac distributions at the ends of the wire, with a temperature higher than the cryostat temperature due to electron heating in the reservoirs \cite{HennySchonenberger,TheseFred,ReservoirNote}.

The link between the two parameters determining the effect of Coulomb interaction, $A$ and $\kappa_{ee}$, can be made explicit by noting that the dephasing rate is the average of the inverse of the lifetime of QPs at energies within $k_{B}T$ of the Fermi energy \cite{AltshulerSimon} :%
\begin{align*}
\frac{1}{\tau_{\varphi}} & \approx 2\int_{\hbar/\tau_{\varphi}}^{k_{B}T}\d\varepsilon\frac{\kappa_{ee}}{\varepsilon^{3/2}}k_{B}T \\
& \approx \frac{4\kappa_{ee}}{\sqrt{\hbar/\tau_{\varphi}}}k_{B}T
\end{align*}
so that
\begin{equation*}
\frac{1}{\tau_{\varphi}}\approx\left( \frac{4\kappa_{ee}k_{B}}{\sqrt{\hbar}}\right) ^{2/3}T^{2/3}.
\end{equation*}
While this derivation reproduces the correct dependence on sample
parameters of the more rigorous theory \cite{AAK,AleinerWav}, the
prefactor depends on the exact value of the cutoff, whose order of magnitude is $\hbar /\tau_{\varphi}$.
The choice of the cutoff can be made so that our derivation stays consistent with the expressions \textsc{Eq.}~(\ref{A}),(\ref{kappa}) of $A$ and $\kappa_{ee}$. Thus it is possible to express $A$ as an intensity $\kappa_{A}$ for Coulomb interaction, using

\begin{equation}
A\equiv\left( \frac{\pi\kappa_{A}k_{B}}{2\sqrt{\hbar}}\right) ^{2/3}. \label{kappaA}
\end{equation}

\section{Comparison between experimental and theoretical results for both methods}

We present here data taken on wires deposited from 6N-purity
(99.9999\%) silver sources. The fabrication procedure for weak localization type (WL)
samples is described in \textsc{Ref.}~\cite{PRB}. The
sample parameters are given in \textsc{Table}~\ref{tableParam}
(weak localization measurements) and \textsc{Table}~\ref{Relax
Param} (energy relaxation measurements). The names of the samples used in energy relaxation (Relax) experiments contain Roman numerals, which indicate the index of the
experiment, and a number, which is the approximate wire length
in microns.  Most Relax samples were
obtained in a single step, using two-angle evaporations through a
suspended mask \cite{TheseFred}. Samples AgII5 and AgII10, on the one
hand, and AgIV20$\alpha$ and AgIV20$\beta$, on the other hand,
were fabricated at the same time, on the same chip. Samples
AgXI10, AgXII40 and AgXV40 were fabricated in two steps of
e-beam lithography: in a first step, the wire pattern was defined,
then silver was evaporated and followed by a lift-off, and a new
deposition of resist. In a second step, the pattern for the
aluminum electrodes was exposed to the electron beam. In the
vacuum chamber of the deposition machine, the silver layer was
cleaned by argon ion milling. A thin (3 nm) layer of aluminum was
then deposited, followed by an oxidation in 1.3 mbar of
oxygen-argon (20\%-80\%) during 8 minutes, in order to form the
tunnel barrier. Finally, a layer of aluminum was deposited. 

\begin{table}[b]
\begin{center}
\begin{tabular}{|c|c c c c c|}\hline
Sample & $L$ & $w$ & $t$ & $R$ & $D$\\
        & ($\mu$m) & (nm) & (nm) & (k$\Omega$)& ($\mathrm{cm}^{2}/\mathrm{s}$)\\\hline
Ag(6N)a & 136 & 65 & 47 & 1.44 & 117\\
Ag(6N)b & 271 & 100 & 45 & 3.30 & 69.2\\
Ag(6N)c & 400 & 105 & 53.5 & 1.44 & 187\\
Ag(6N)d & 285 & 90 & 36 & 2.00 & 167\\\hline
\end{tabular}
\vspace{0.15cm}
\caption{Geometrical and electrical characteristics of samples for weak localization measurements.  The diffusion coefficient $D$ is obtained using Einstein's relation $1/\rho=\nu_{F}e^{2}D$ with the density of states in silver $\nu_{F}=1.03\times10^{47}~\mathrm{J^{-1}m^{-3}}$, and the resistivity $\rho$ extracted from the resistance $R$, thickness $t$, length $L$ and width $w$ of the wire.} \label{tableParam}
\end{center}
\end{table}

\begin{figure}[ptbh]
\begin{center}
\includegraphics[width=3in]{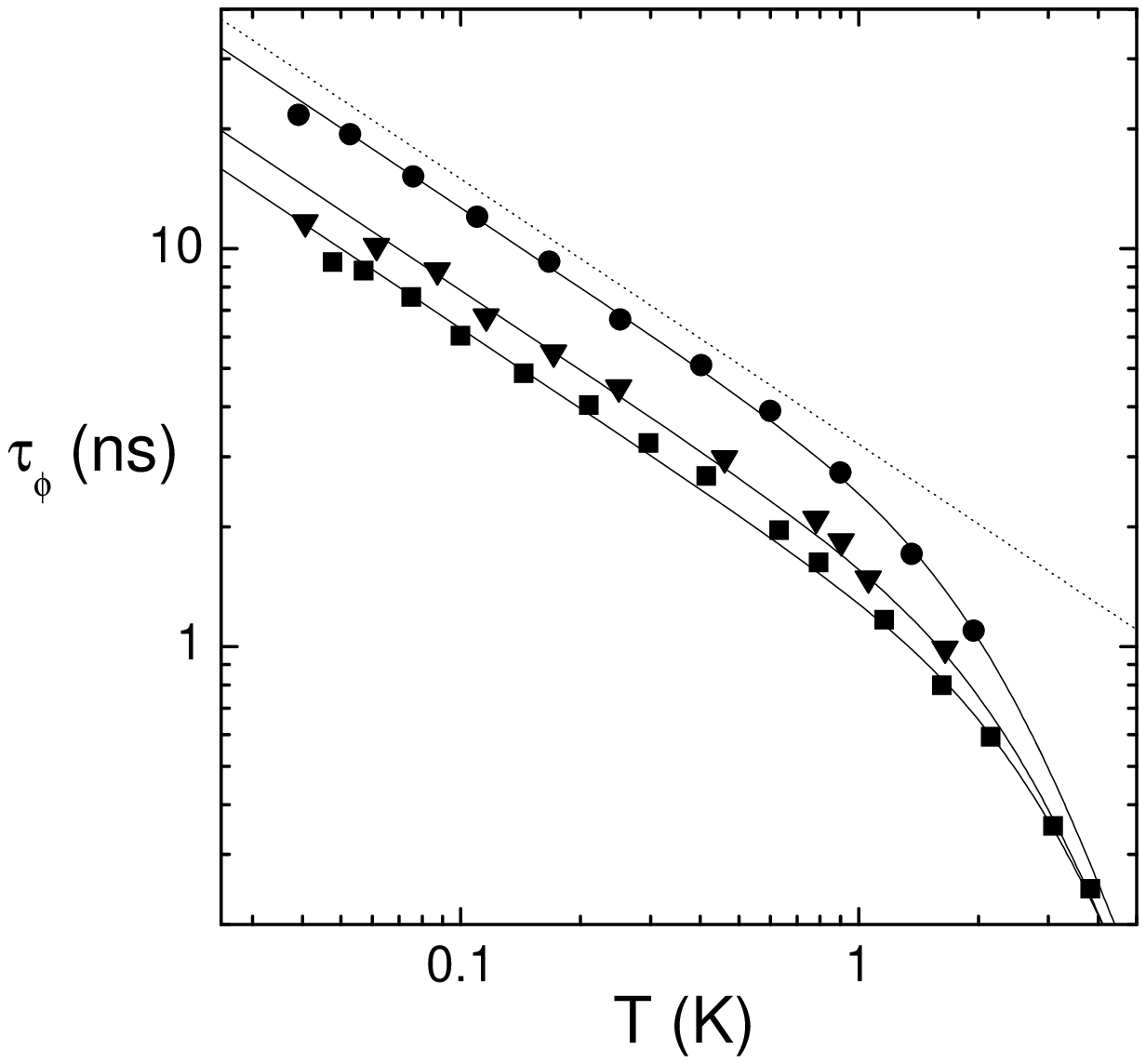} \caption{Phase coherence time vs
temperature in samples Ag(6N)a ($\scriptstyle{\blacksquare}$), Ag(6N)b ($\blacktriangledown$), and Ag(6N)c ($\bullet$), all made of 6N sources. Continuous lines are fits of the data to \textsc{Eq.}~(\ref{tauphi}).  The quantitative prediction of \textsc{Eq.}~(\ref{A}) for electron-electron interactions in sample Ag(6N)c is shown as a dashed line.} \label{Ag6n}
\end{center}
\end{figure}

\begin{table}[b]
\begin{center}
\begin{tabular}{|c| c @{           } c c |}\hline
Sample & $A^{\textrm{thy}}$ & $A$ & $B$\\
$\blacksquare$&  \multicolumn{2}{c}{(ns$^{-1}$K$^{-2/3}$)} & (ns$^{-1}$K$^{-3}$)\\\hline
Ag(6N)a & 0.55 & 0.73 & 0.045\\
Ag(6N)b & 0.51 & 0.59 & 0.05\\
Ag(6N)c & 0.31 & 0.37 & 0.047\\
Ag(6N)d & 0.47 & 0.56 & 0.044\\\hline
\end{tabular}
\vspace{0.15cm}
\caption{Theoretical predictions of \textsc{Eq.}~(\ref{A}) ($A^{\textrm{thy}}$) and fit parameters ($A$ and $B$) for $\tau_{\varphi}(T)$ in the  samples of Table 1 using the functional form given by \textsc{Eq.}~(\ref{tauphi}). Comparison of $A^{\textrm{thy}}$ and $A$ is shown graphically in \textsc{Fig.}~\ref{Comparison}.}
\label{tableFitEE}
\end{center}
\end{table}

\begin{table}[hbtp]
\begin{center}
\begin{tabular}{|c|c c c c c c|}\hline
Sample &            $L$ &       $w$ & $t$  & $R$ &  $D$ & $\tau_D$ \\
    & ($\mu$m) & (nm) & (nm) & ($\Omega$) & ($\mathrm{cm}^{2}/\mathrm{s}$) & (ns) 
\\\hline
AgI5 &              5.05 &      90 & 43 &  41 &  121 & 2.1 \\
AgII5 &             5.2 &       66 & 39 &  44 & 173 & 1.6 \\
AgII10 &            10.3 &      65 & 39 &  81 & 191 & 5.6 \\
AgIII20 &           19.6 &      160 & 43 &  45 & 241 & 16  \\
AgIV20$\alpha$ &    19.7 &      95 & 44 &  86 & 208 & 19 \\
AgIV20$\beta$ &     19.9 &      100 & 44 &  91 & 188 & 21 \\
AgX20 &              21.7 &     100 & 48 &  80 & 214 & 22 \\
AgXI10 &            9.55 &      124 & 45 &  31 & 211 & 43 \\
AgXII40 &            38 &      180 & 45 &  108(\cite{R not measured}) & 165  & 87\\
AgXV40 &            38 &      145 & 45 &  134 & 165 & 87\\\hline
\end{tabular}
\vspace{0.15cm}
\caption{Geometrical and electrical characteristics of samples for energy relaxation measurements.} \label{Relax Param}
\end{center}
\end{table}

\begin{figure}[bpth]
\begin{center}
\includegraphics[width=3in]{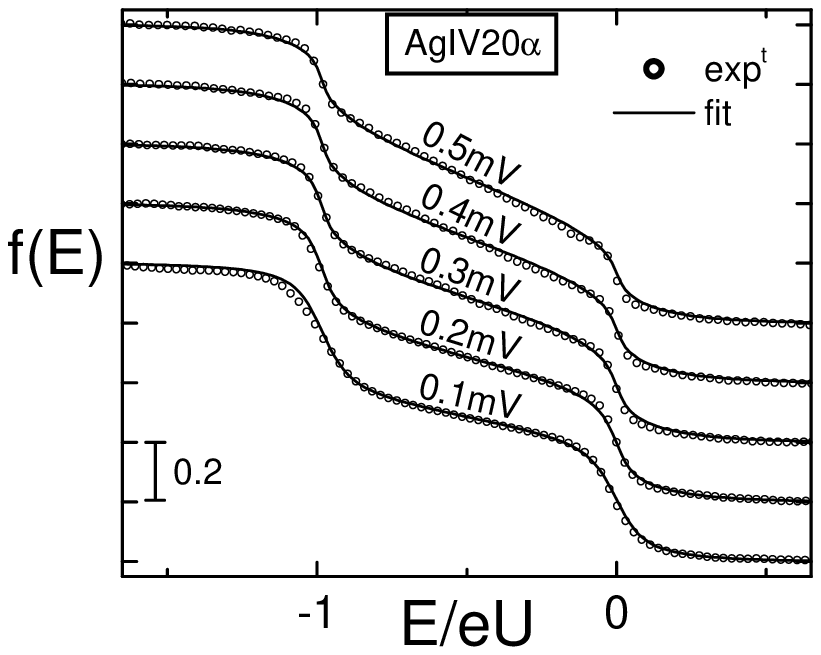} \caption{Measurements ($\circ$) and fits (solid curves) of the quasiparticle energy distribution function $f_{\frac{1}{2}}(E)$ for five different values of the applied voltage $U$ across the wire AgIV20$\alpha$. The data have been shifted vertically for clarity.} \label{AgRelaxstd}
\end{center}
\end{figure}

\begin{table}[hbtp]
\begin{center}
\begin{tabular}{|c| c c c c |}\hline
Sample & &  $\kappa^{\textrm{thy}}_{ee}$ & $\kappa_{ee}$ & \\
$\circ$ & & \multicolumn{2}{c}{(ns$^{-1}$meV$^{-1/2}$)} & \\\hline
AgI5 &     &          0.060 &         $0.95$ & \\
AgII5 &     &        0.076 &       $0.5$  & \\
AgII10 &     &       0.073 &      $0.54$  & \\
AgIII20 &     &      0.024 &      $0.5$    &\\
AgIV20$\alpha$ & &    0.043 &      $0.40$  & \\
AgIV20$\beta$ & &     0.043 &      $0.37$   & \\
AgX20   &       &    0.037 &       $0.11$  & \\
AgXI10 &        &    0.032 &         $< 0.18$&\\
AgXII40 &       &    0.025 &      $0.18$  &\\
AgXV40 &      &    0.031 &         $0.32$&\\
\hline
\end{tabular}
\vspace{0.15cm}
\caption{Theoretical predictions of \textsc{Eq.}~(\ref{kappa}) ($\kappa^{\textrm{thy}}_{ee}$) and fit parameters ($\kappa_{ee}$) for $f_x(E)$ in the  samples of Table 3 using the solution of the Boltzmann equation \textsc{Eq.}~(\ref{Boltzmann}). The distribution functions measured on sample AgXI10 were so close to the noninteracting regime that it was only possible to give an upper bound to the value of $\kappa_{ee}$. Comparison of $\kappa^{\textrm{thy}}_{ee}$ and $\kappa_{ee}$ is shown graphically in \textsc{Fig.}~\ref{Comparison}.}\label{tableauFitBoltz}
\end{center}
\end{table}

\begin{figure}[ptbh]
\begin{center}
\includegraphics[width=3in]{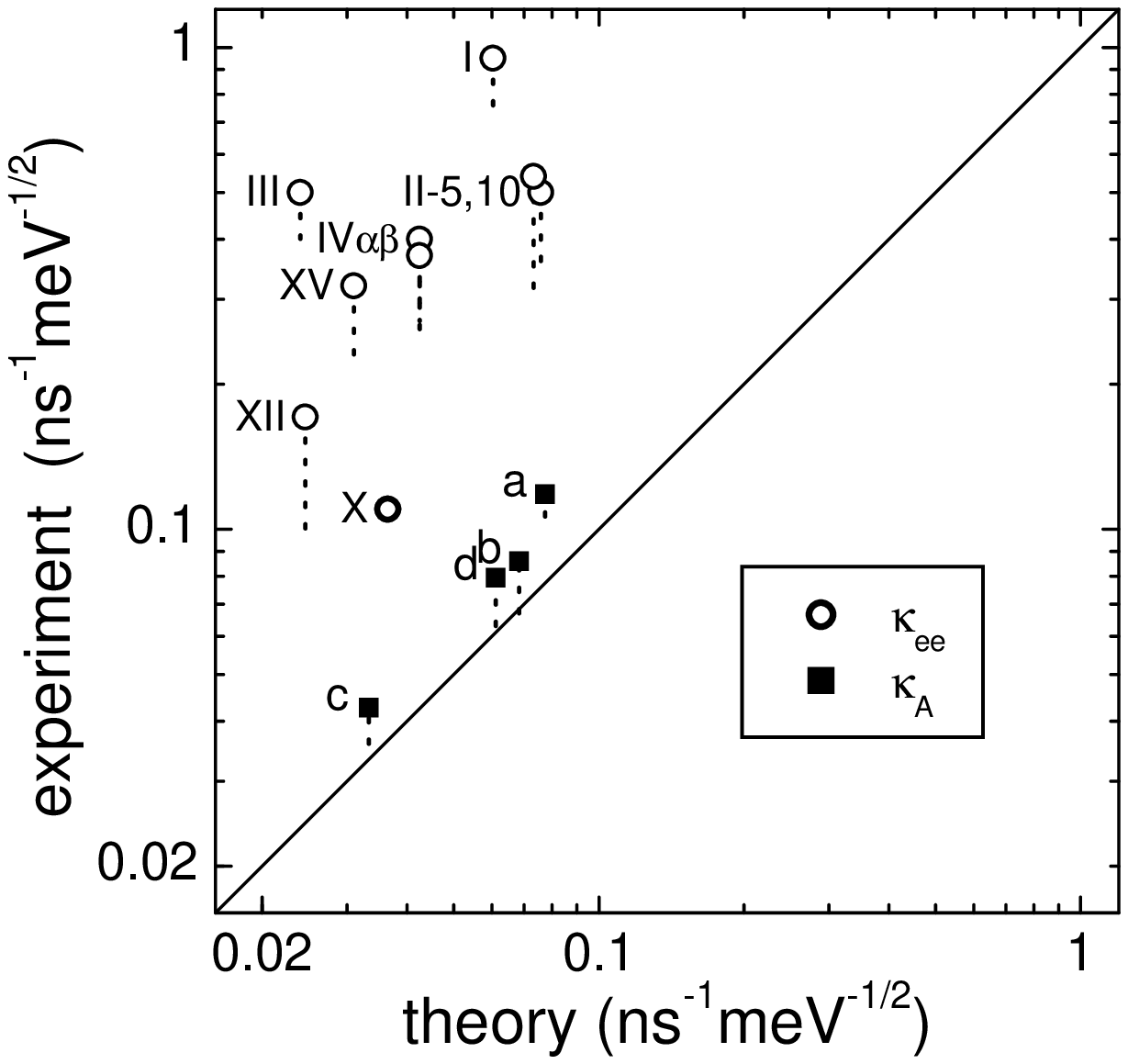} \caption{Comparison of the experimental prefactor with the theoretical prediction \textsc{Eqs.}~(\ref{A}),(\ref{kappa}),(\ref{kappaA}), for weak localization experiments ($\scriptstyle{\blacksquare}$) and energy relaxation experiments ($\circ$). If we assume that a small amount of magnetic impurities is present in the WL samples, the fit values of $\kappa_A$ can be reduced down to the bottom of the dashed lines below the squares. Similarly for the Relax experiments, if we assume that magnetic impurities are present, we obtain a range of values of $\kappa_{ee}$ compatible with the data, represented as a dashed line below the $\circ$. The behavior of sample AgX20 was measured in a magnetic field, allowing us to place an upper bound on the concentration of magnetic impurities, and hence to rule out the possibility of reducing the value of $\kappa_{ee}$ more than 15\%. Thus, this data point is represented as a bold circle without any dashed line. \label{Comparison}}
\end{center}
\end{figure}

In \textsc{Fig.}~\ref{Ag6n}, we present
$\tau_{\varphi}(T)$ for the first three WL samples (the data points
of the last one, which are presented in Ref. \cite{PRB}, are so
close to those of the third one that they would confuse the figure), as
well as the best fits with \textsc{Eq.}~(\ref{tauphi}). The fit
parameters are given in \textsc{Table}~\ref{tableFitEE}. The fit
value of $A$ is very close to the theoretical value for the
exchange contribution of the Coulomb interaction, as can be seen
in \textsc{Fig.}~\ref{Comparison} where the X-coordinate of the
solid squares is the theoretical value of $\kappa_{A}$ using
\textsc{Eqs.}~(\ref{A}) and (\ref{kappaA}), and the Y-coordinate
is the value from experiment.

The situation is quite different in energy relaxation experiments.
We show in \textsc{Fig.}~\ref{AgRelaxstd} distribution functions $f(E)$ measured in the middle of sample AgIV20$\alpha$, for $U$ ranging from 0.1 to 0.5~mV, plotted as a function of the reduced energy $E/eU.$ Solid lines are fits resulting from the numerical solution of the Boltzmann equation, obtained with $\kappa_{ee}=0.40~$ns$^{-1}$meV$^{-1/2}$. The increase in slope of the middle step of $f(E)$ when $U$ increases, characteristic of the effect of Coulomb interaction, is well reproduced. However, the fit value for $\kappa_{ee}$ is nearly an order of magnitude larger than the value given by \textsc{Eq.}~(\ref{kappa}). Similar discrepancies exist for the other Relax samples.
It could be argued that the numerical prefactor in
\textsc{Eq.}~(\ref{kappa}) is incorrect. \textsc{Fig.}~\ref{Comparison} seems to rule out this explanation: the circles corresponding to the theoretical and fit values, given also in \textsc{Table}
\ref{tableauFitBoltz}, present a large scatter,
and so the ratio between experiment and theory does not appear to be constant.

\section{Discussion of the discrepancy between the two experiments}

Figure \ref{Comparison} reveals a very puzzling difference between
weak localization (WL) and energy relaxation (Relax) experiments.
Whereas the results of both types of experiments are precisely
accounted for by the theory of Coulomb interactions in disordered
wires as far as the energy dependence is concerned, the prefactor
is well understood for the first, but not at all for the second.
In order to resolve this puzzle, we now list the differences
between the two types of experiments:

\subsection{Possibility of extrinsic energy exchange processes in Relax samples}

\label{extrinsic}

WL experiments are extremely sensitive to very small quantities of magnetic impurities.  It was shown in \cite{PRB} that even in our cleanest Ag(6N) wires, there was evidence for magnetic impurities at concentrations of about 0.01~ppm, i.e. 1 impurity atom for every $10^8$ Ag atoms.  Their contribution to $\tau_{\varphi}$ was visible only at the lowest experimental temperatures. In \textsc{Fig.}~\ref{Comparison}, we have indicated with the vertical dashed lines how far the fit values of $\kappa_A$ can be reduced if one includes a small concentration of magnetic impurities as an extra fit parameter.

It is now undesrtood that magnetic impurities also mediate energy exchange between electrons \cite{KG,PRLAnne}.  Could the presence of magnetic impurities explain the anomalously large apparent values of $\kappa_{ee}$ observed in many Relax experiments? Since most of the Ag samples used in the WL experiments were fabricated in the same deposition system used for the Relax samples, we expect that Relax samples should be equally clean. This hypothesis must be checked, however. The presence of magnetic impurities in Relax samples can be detected directly by performing the experiment as a function of magnetic field \cite{PRLAnne}. In samples AgX20 and AgXI10, the magnetic field dependence of the measurements set an upper bound to the concentrations of magnetic impurities at 0.1 and 0.6~ppm respectively. For sample AgX20, if we include the effect of 0.1~ppm of magnetic impurities into the analysis of the Relax data, the value of $\kappa_{ee}$ is reduced by only 15\%.  In sample AgXI10, the distribution functions were so close to the noninteracting regime that it was only possible to place an upper bound on $\kappa_{ee}$, hence this sample does not appear in \textsc{Fig.}~\ref{Comparison}.

For the Relax samples that were not measured in a magnetic field, no upper bound to the concentration of magnetic impurities is experimentally determined. We have estimated the resulting systematic uncertainty in $\kappa_{ee}$ by the following analysis.  We have assumed that electron-electron interactions mediated by magnetic impurities contribute to energy exchange.  For this process, the interaction kernel is approximately $K(\varepsilon) = \kappa_{2}\varepsilon^{-2}$ \cite{KG,GG}. If we fit the data using the value of $\kappa_{2}$ as an additional fit parameter, we can ask how small the value of $\kappa_{ee}$ can become before the fits become clearly incompatible with the data. The results are shown by the dashed lines descending below the points for the Relax samples in \textsc{Fig.}~\ref{Comparison}.  As can be seen, for some samples the fits are somewhat insensitive to the relative weights of $\kappa_{ee}$ and $\kappa_2$, and the discrepancy between theory and experiment gets smaller. Nevertheless, the discrepancy still remains.  We conclude for the time being that extrinsic energy exchange processes with $K(\varepsilon) \propto \varepsilon^{-2}$ are unlikely to explain completely the discrepancy between experiment and theory.  This issue will be discussed further in section~\ref{conclusion}.

\subsection{Sample dimensionality}

\begin{figure}[ptbh]
\begin{center}
\includegraphics[width=3in]{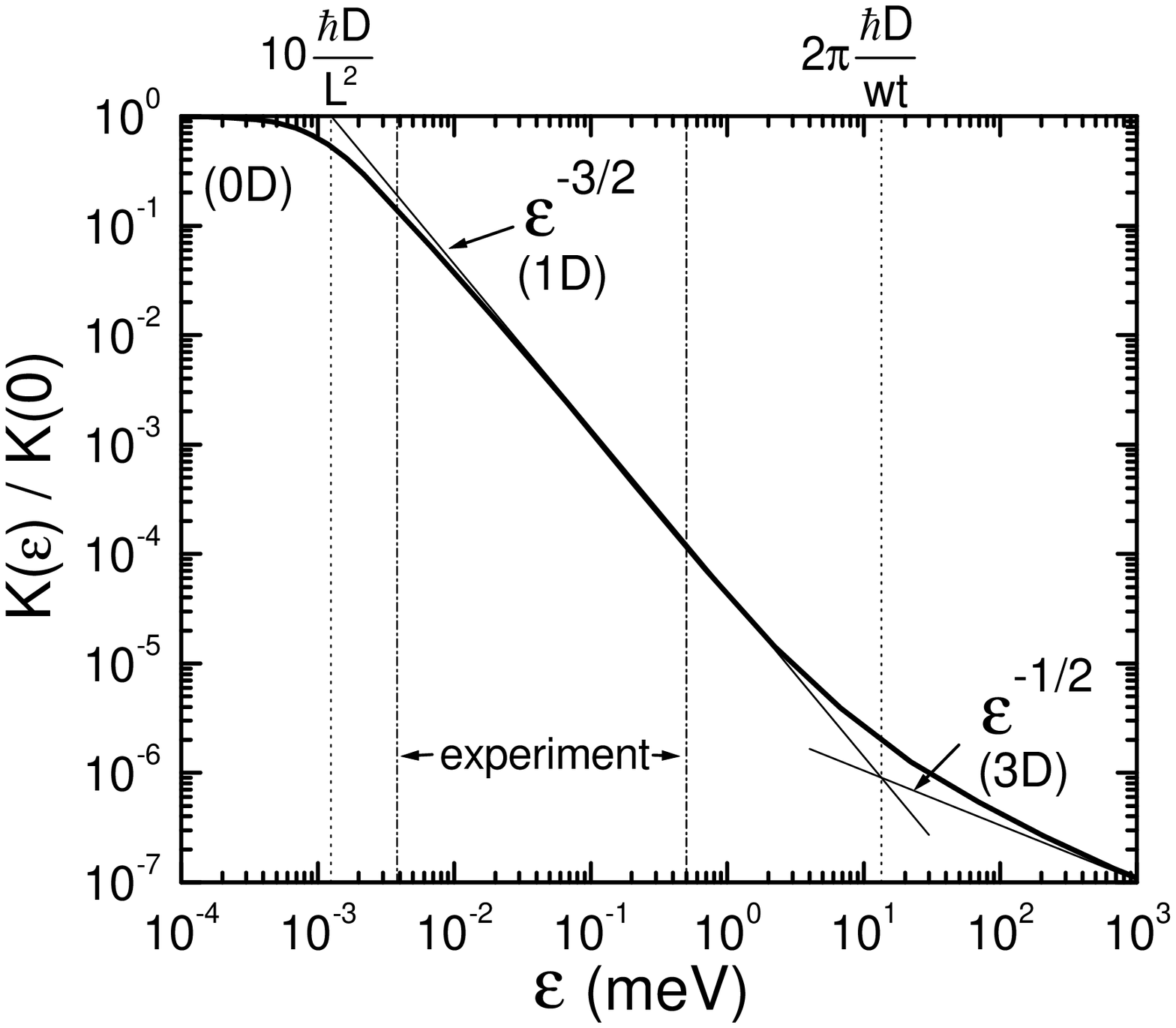} \caption{Energy dependence of the kernel $K(\varepsilon)$ of Coulomb interaction in a wire with $L=10~\mu\textrm{m},w=130~\textrm{nm},t=45~\textrm{nm}$ and $D=200~\textrm{cm}^2/\textrm{s}$. The asymptotic zero-, one- and three-dimensional regimes (0D, 1D, 3D) are characterized by $K(\varepsilon)=K(0)$, $K(\varepsilon)\propto \varepsilon^{-3/2}$ and $K(\varepsilon)\propto \varepsilon^{-1/2}$, respectively (straight lines). The two-dimensional regime is not clearly visible because $w \approx t.$ The range of relevant $\varepsilon$'s for the Relax experiments is determined by $k_BT_{\textrm{min}}$ and $eU_{\textrm{max}}$. The normalization factor on the y-axis is $K(0)= \left(45\pi(\hbar D/L^2)^2\hbar\nu_F wtL\right)^{-1}$\label{K(E)}}.
\end{center}
\end{figure}

The intensity and energy dependence of Coulomb interaction depends on sample dimensionality \cite{AA}. The one-dimensional (1D) regime described in section 2 corresponds, in WL experiments, to situations where $w,t\ll L_{\varphi}\ll L$. This inequality is well obeyed in our experiments, where $L_{\varphi}$ varies between $1~\mu$m to $20~\mu$m. In practice, the wire length $L$ was chosen much larger than $L_{\varphi}\left( T_{\min}\right)$, where $T_{\min}$ is the lowest experimental temperature, in order to reduce the amplitude of conductance fluctuations, which spoil the analysis of the magnetoresistance in terms of the WL theory.\\
In Relax experiments, on the other hand, the distribution function $f(E)$ only contains information on
the interaction process if it is far from a Fermi function and far from a perfect double-step, \emph{i.e. }if $L\approx \textrm{few }L_{\varphi}\left( eU_{\max}/k_{B}\right)$. Thus the
wire length is smaller than for the WL experiments. The dimensionality criterion for Relax is illustrated in \textsc{Fig.}~\ref{K(E)}, where we plot the function $K(\varepsilon)$
calculated using the discrete sum over the longitudinal and transverse wave vectors \cite{Gilles,Blanter}

\begin{equation}
K(\varepsilon)\propto \sum_{\substack{q_x\neq 0 \\ q_y,q_z}}\frac{1}{D^2\vec{q}^4+\left(\varepsilon/\hbar\right)^2} \label{K(eps)}
\end{equation}
where $q_x=\frac{\pi n_x}{L}$, $q_y=\frac{\pi n_y}{w}$ and $q_z=\frac{\pi n_z}{t}$ are the wave vector components with $n_x\in\Nset^*$ and $n_y,n_z\in\Nset$.

Typical sample dimensions were chosen: $L=10~\mu\textrm{m},$ $w=130~\textrm{nm},$ $t=45~\textrm{nm}$ and $D=200~\textrm{cm}^2/\textrm{s}$. \textsc{Fig.} \ref{K(E)} shows that for all relevant energies in the experiments, $K(\varepsilon)$ is far from the 1D-3D transition. For small energies near $k_BT_{min}$, the behavior of $K(\varepsilon)$ differs slightly from the one-dimensional $\varepsilon^{-3/2}$ power law, but this deviation goes in the wrong direction to explain the discrepancy between theory and experiment.

\subsection{Diffusive approximation in narrow wires}

The energy scales probed by WL and Relax experiments are rather different.  In wires, the value of $\tau_{\varphi}$ is essentially determined by the low energy cut-off of the interaction, at $\hbar/\tau_{\varphi}.$ In the samples presented here, $\tau_{\varphi}$ ranges (in the relevant temperature range: 1~K down to 40~mK) from 1 to 20~ns, corresponding to energies $\hbar/\tau_{\varphi}$ between 0.03 and 0.6~$\mu$eV. In the Relax experiments, the shape of $f(E)$ is entirely determined by energy exchanges of an amount between $k_BT$ and $eU$, in practice between 4 and 500~$\mu$eV. According to \textsc{Eq.}~(\ref{K(eps)}), the characteristic lengthscale $1/q=\sqrt{\hbar D/\varepsilon}$ for the interaction is therefore a few micrometers for WL, several hundreds of nanometers for Relax. The discrepancy between the results of the two types of experiment could point out a failure of the diffusive model, in which the QP dynamics is described by a single diffusion constant $D$. This argument is reinforced by the fact that the elastic mean free path deduced from $D$ is of the order of the wire thickness $t$, indicating that surface and grain boundary scattering dominate the elastic processes. If surface scattering alone were dominant, the elastic mean free path of QPs with an instantaneous wavevector along the axis of the wire would be very different from that of QPs travelling in a perpendicular direction, and the diffusive approximation would break down. To our knowledge, Coulomb interaction has never been investigated in this regime. However it is not clear why this situation could be described by the same energy dependence and why the intensity could be larger.

\subsection{Departure from equilibrium}

WL experiments are performed very close to equilibrium.  In Relax experiments, a voltage $U\gg k_{B}T/e$ is applied to the wires in order to establish an out-of-equilibrium situation. Near the Fermi level, the distribution function is very different from a Fermi function, and it could be argued that the derivation leading to the expression (\ref{kappa}) of the prefactor $\kappa_{ee}$ is no longer valid. 
In order to test this hypothesis, we have performed a complementary experiment, described below, in which the effect of the distance to equilibrium is investigated.

\section{A new Relax experiment close to equilibrium}

\begin{figure}[ptbh]
\begin{center}
\includegraphics[width=3.0in]{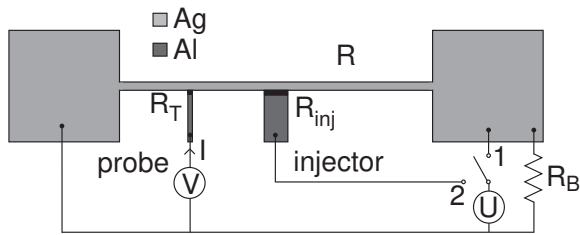} \caption{Schematic diagram of an experiment to measure $f_{x}(E)$ in a wire close to equilibrium.  Quasiparticles are injected into the wire from a superconducting wire (labelled \emph{injector}) through a tunnel junction biased at potential $U$ (switch position 2).  The distribution function $f_{x}(E)$ at position $x=0.25$ is then determined from the $\d I/\d V$ characteristic of the probe junction. Alternatively, the wire can be driven far from equilibrium by applying the voltage bias $U$ across the wire (switch position 1). The resistance $R_B$ is chosen so that the potential of the right reservoir remains close to zero when the switch is in position 2.}\label{INJ scheme}
\end{center}
\end{figure}

\begin{figure}[ptbh]
\begin{center}
\includegraphics[width=3.0in]{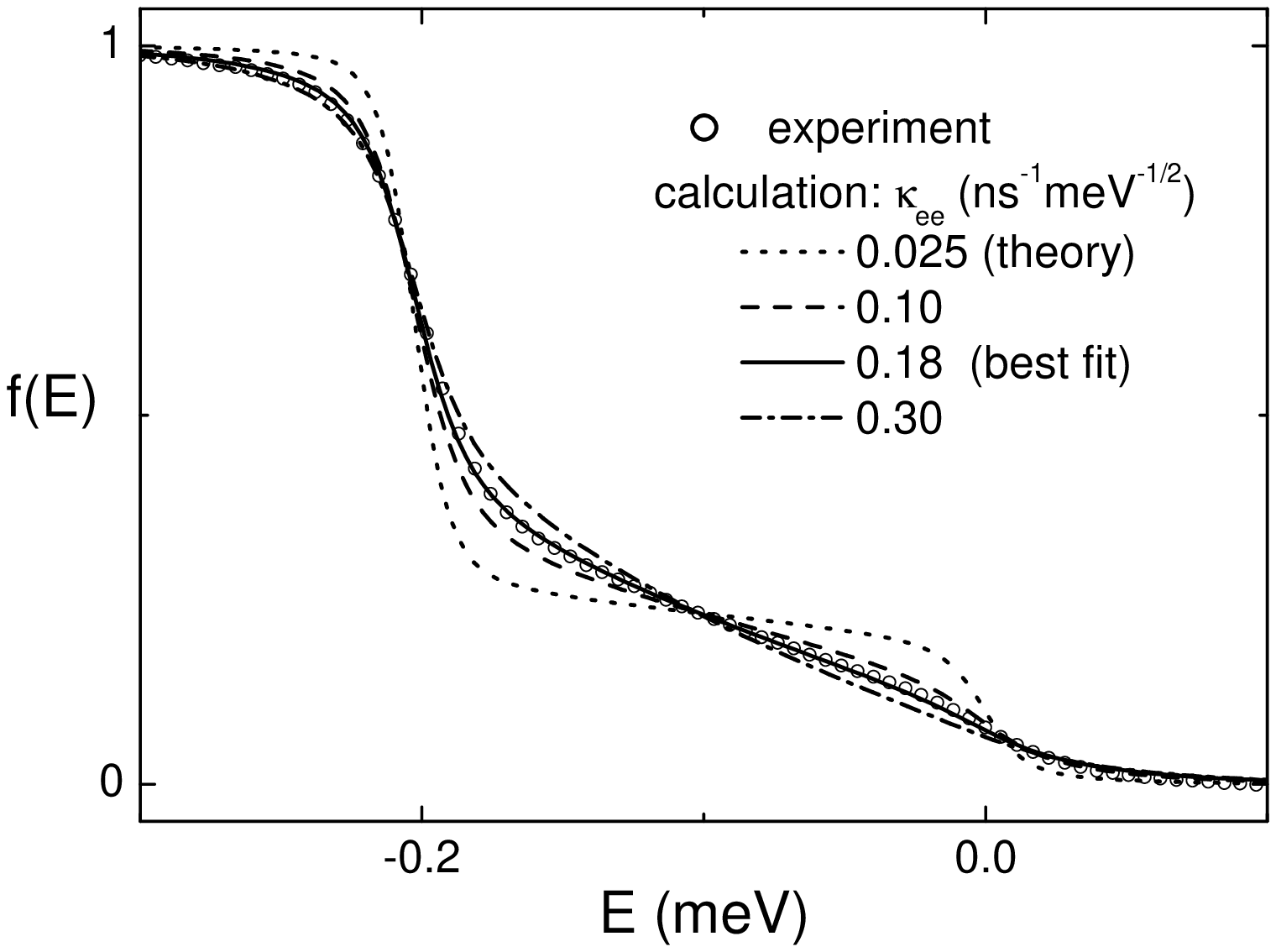} \caption{Measured ($\circ$) distribution function $f_{\frac{1}{4}}(E)$ in the ``conventional'' Relax experiment using sample AgXII40 with the switch of \textsc{Fig.}~\ref{INJ scheme} in position 1, and for $U=0.2~$mV.  The solid line is a numerical solution to the Boltzmann equation using the prefactor $\kappa_{ee}=0.18~\mathrm{ns^{-1}meV^{-1/2}}$ for the Coulomb interaction between electrons.  As shown by the three dot-dashed lines, other values of $\kappa_{ee}$ produce markedly worse fits to the data. In particular, the theoretical value $\kappa_{ee}=0.025~\textrm{ns}^{-1}\textrm{meV}^{-1/2}$ does not come close to reproducing the experimental results.}\label{INJ cas standard}
\end{center}
\end{figure}

\begin{figure}[ptbh]
\begin{center}
\includegraphics[width=3.0in]{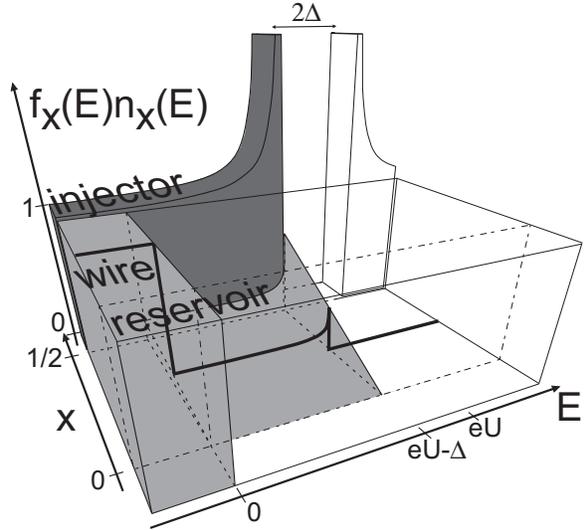} \caption{Schematic diagram showing the spatial and energy dependence of the distribution function $f_{x}(E)$ of QPs driven out-of-equilibrium by the voltage $U$ using the geometry with the switch of \textsc{Fig.}~\ref{INJ scheme} in position 2 (we have assumed $U<-\Delta$). The surrounding box shows the density of states along the circuit and the gray volume shows the occupied states whose normalized density is $f_x(E)n_x(E)$. The inelastic processes involving QPs are assumed to be very weak for clarity. The thick line shows the distribution function $f_{\frac{1}{4}}(E)$ at $x=1/4$.}\label{nf cas eq}
\end{center}
\end{figure}

\textsc{Fig.}~\ref{INJ scheme} shows a schematic of sample
AgXII40, which was designed to investigate the effect of the deviation of $f(E)$ from an equilibrium Fermi distribution. As in other Relax
experiments, a wire (38~$\mu$m long, 180~nm wide, 45~nm thick) is
placed between large contact pads. A superconducting probe
electrode is placed at $x=1/4,$ with a tunnel resistance to the
wire of $15~k\Omega.$ The size of the tunnel junction was
0.18$\times$0.23~$\mu$m$^{2}$. When the switch on
\textsc{Fig.}~\ref{INJ scheme} is placed in position 1, the
``conventional'' Relax experiment can be performed. A measured
distribution function is shown in \textsc{Fig.}~\ref{INJ cas
standard}. The intensity of the Coulomb interaction deduced from
the fits of $f(E)$ is $\kappa_{ee}=0.18~$ns$^{-1}$meV$^{-1/2}$, as
indicated in \textsc{Table}~\ref{tableauFitBoltz}.
\textsc{Eq.}~(\ref{kappa}) has been used \cite{R not measured} to
calculate the theoretical value
$\kappa_{ee}^{\textrm{thy}}=0.025~$ns$^{-1}$meV$^{-1/2}$. This discrepancy
is of the same type as the one observed in the other samples of
\textsc{Table}~\ref{tableauFitBoltz}. A second superconducting
electrode, denoted \emph{injector} in \textsc{Fig.}~\ref{INJ
scheme}, forms a tunnel junction with the wire around its center,
but with a much smaller resistance $R_{\textrm{inj}}=1.1~\textrm{k}\Omega$ than the probe junction, resulting from a larger area:
0.57$\times 0.8~\mu\textrm{m}^{2}$. This junction was obtained at the
overlap between the $w_{\textrm{inj}}=0.8~\mu$m-wide
superconducting electrode and the wire, which presents an
intentional broadening at this position. When the switch of
\textsc{Fig.}~\ref{INJ  scheme} is placed in position 2,
quasiparticles are injected through the tunnel junction into the
wire when $|U|>\Delta/e,$ with $\Delta$ the gap in the QP density of
states of the injector. On the normal side of the tunnel junction,
the QP distribution function is therefore expected to display a
step, the shape of which reflects the BCS density of states
$n_{S}(E)=\textrm{Re}\left(\left| E\right| /\sqrt{E^{2}-\Delta^{2}}\right)$. The height of
the step away from the BCS peak is given by the ratio of the
injection rate of QPs to the diffusion rate towards the two normal
reservoirs:
$f_{\frac{1}{2}}(E)\sim\left(\frac{R}{4}\right)/R_{\textrm{inj}}\equiv r$ (the
factor 1/4 results from the parallel combination of the two halves
of the normal wire as will be shown below). A quantitative
description follows from the introduction of new boundary
conditions in the Boltzmann \textsc{Eq.}~(\ref{Boltzmann}): $f_x(E)$
is a Fermi function with a zero electrochemical potential at $x=0$
and $-eU_{r}$ at $x=1$, whereas at $x=\frac{1}{2}$ current
conservation at each energy implies
\begin{equation*}
\nu_{F}wteD\left( \frac{\partial f_x(E)}{L\partial x}|_{x=\frac{1}{2}^{+}}-\frac{\partial f_x(E)}{L\partial x}|_{x=\frac{1}{2}^{-}}\right) =i_{\textrm{inj}}(E)
\end{equation*}
with
\begin{equation*}
i_{\textrm{inj}}(E)=\frac{1}{eR_{\textrm{inj}}}n_{S}(E+eU) ( f_{S}(E+eU)-f_{\frac{1}{2}}(E))
\end{equation*}
where $f_S(E)$ is the distribution function in the superconducting injector. We neglect here the slight modification of the DOS in the wire due to proximity effect, because of the small transparency of the tunnel barrier.
Finally,
\begin{align}
& \frac{\partial f_x(E)}{\partial x}|_{x=\frac{1}{2}^{+}}-\frac{\partial f_x(E)}{%
\partial x}|_{x=\frac{1}{2}^{-}}  \notag \\
& =\frac{R}{R_{\textrm{inj}}}n_{S}(E+eU)( f_{S}(E+eU)-f_{\frac{1}{2}}(E)) .  \label{CL}
\end{align}
The electrical potential of the right reservoir, which is connected to ground by a bias resistance $R_B=12~\Omega$, is given by $U_{r}=\frac{1}{2}\frac{R R_B}{R+R_B}\int i_{\textrm{inj}}(E)\mathrm{d}E<\frac{R_B}{2R_{\textrm{inj}}}U.$
Since $\frac{R_B}{2R_{\textrm{inj}}}\simeq 0.005$, we make the approximation $U_{r}=0$, so that the situation is symmetric: $f_{x}(E)=f_{1-x}(E)$ and \textsc{Eq.}~(\ref{CL}) becomes
\begin{align}
& \frac{\partial f_x(E)}{\partial x}|_{x=\frac{1}{2}^{+}}  =-\frac{\partial f_x(E)}{\partial x}|_{x=\frac{1}{2}^{-}}  \notag \\
& =2r n_{S}(E+eU)( f_{S}(E+eU)-f_{\frac{1}{2}}(E)) . \label{CL2}
\end{align}
In the absence of interactions, at $T=0,$ one obtains directly for $x<\frac{1}{2}$ (assuming $U<-\Delta$):
\begin{equation*}
f_x(E) = \left\{ \begin{array}{l l}
1 & \text{ for }E<0 \\
2xf_{\frac{1}{2}}(E) & \text{ for }E\in\left[ 0,-eU-\Delta\right] \\
0 & \text{ for }E>-eU-\Delta \end{array}\right.
\end{equation*}
and
\begin{equation}
f_{\frac{1}{2}}(E)=\frac{r~n_{S}\left( E+eU\right) }{1+rn_{S}\left( E+eU\right) }.  \label{BC}
\end{equation}
The spatial dependence of $f_{x}(E)$ is plotted in \textsc{Fig.}~\ref{nf cas eq} for $x<\frac{1}{2}$, assuming $r=0.1$ for visibility (in the experiment, $r \simeq 0.025$). It is seen that $f_{x}(E)$ is much closer to a Fermi function than when the voltage is applied across the wire.

\begin{figure}[ptbh]
\begin{center}
\includegraphics[width=3.0in]{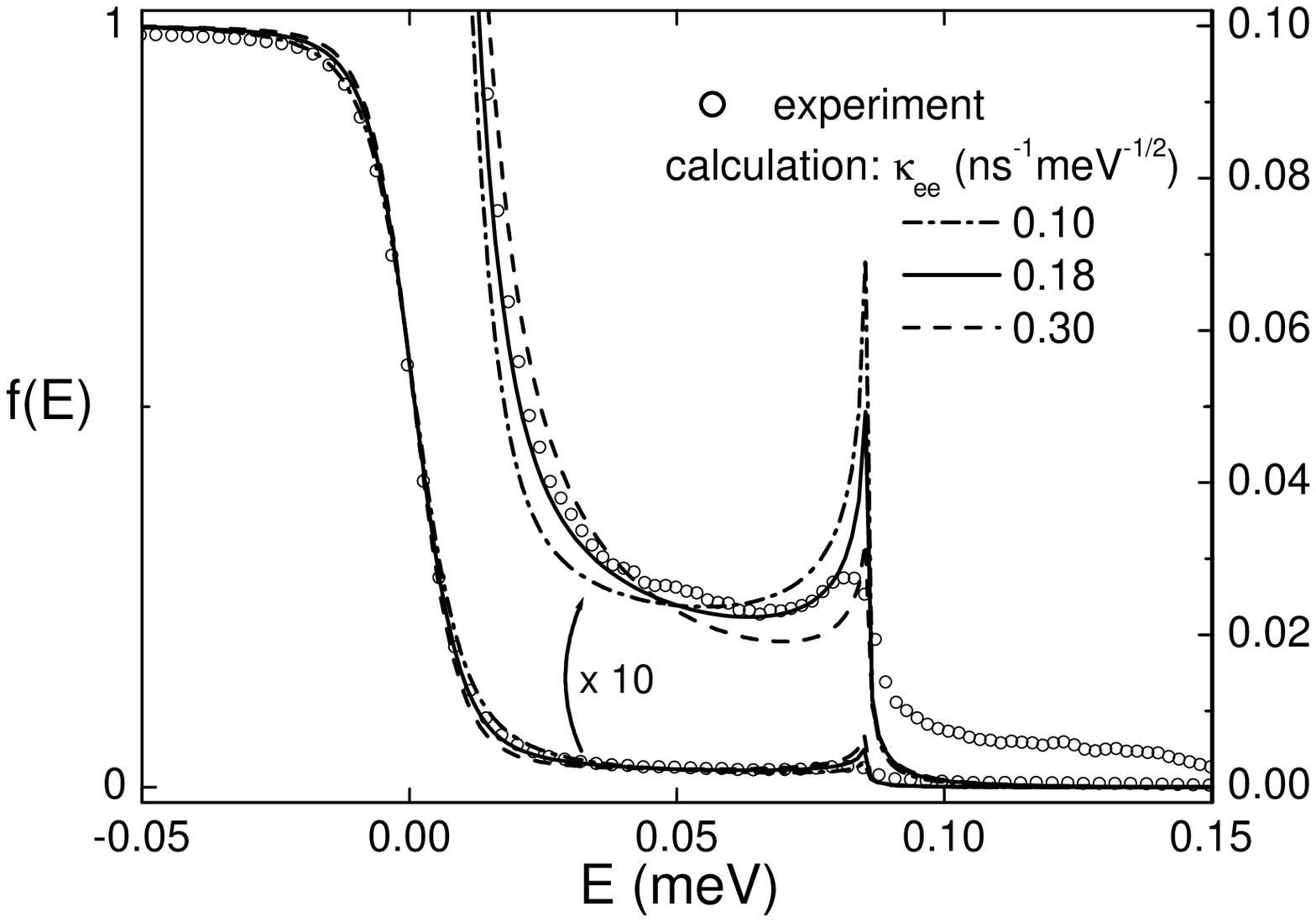} \caption{Measured ($\circ$) distribution function $f_{\frac{1}{4}}(E)$ in the new Relax experiment using sample AgXII40 depicted in \textsc{Fig.}~\ref{INJ  scheme}, with the switch in position 2, and for $U=-0.27~$mV.  The data are also shown magnified by a factor 10 (right scale).  The solid line is a numerical solution to the Boltzmann equation with boundary condition given by \textsc{Eq.}~(\ref{BC}), using as prefactor for the Coulomb interaction $\kappa_{ee}=0.18~\mathrm{ns^{-1}meV^{-1/2}}$.  The two dot-dashed lines show that other values of $\kappa_{ee}$ produce markedly worse fits to the data.}\label{INJ cas eq}
\end{center}
\end{figure}

An experimental curve, obtained for $U=-0.27~$mV, is shown in
\textsc{Fig.}~\ref{INJ cas eq}. As predicted, it presents a very
small step  $(f_{\frac{1}{4}}(E)\approx 0.025)$ extending from $E=0$ to $E=-eU-\Delta,$ with
$\Delta=0.18~$mV the gap for the injector deduced from its $I-V$
characteristic, measured separately. The blow-up ($\times10,$
right scale) shows the expected small peak near $E=-eU-\Delta$. We also show
$f(E)$ calculated using the same parameters as those deduced from
the ``conventional'' measurement, using
\textsc{Eq.}~(\ref{Boltzmann}) and (\ref{CL2}). Except for a
slight rounding of the small peak, the agreement is within
experimental accuracy for all the values of $U$ for which data
were taken ($-0.22$ to $-0.31~\textrm{mV}$). We show in particular that other
values of $\kappa_{ee}$ would produce curves which significantly
differ from the measured one.  Hence the value of $\kappa_{ee}$
deduced from energy exchange experiments does not seem to depend
on whether the distribution is far from equilibrium, as in the
original experiment (\textsc{Fig.}~\ref{INJ cas standard}), or
close to equilibrium, as in the newer experiment described here.
Our conclusion is that Coulomb interaction is not modified by the
fact that $f(E)$ is not exactly a Fermi function.

\section{Conclusions}
\label{conclusion}

In Section~\ref{extrinsic}, we discussed the possibility that the anomalously high rates of energy exchange observed in many Relax experiments could be caused by residual magnetic impurities.  Two arguments against this hypothesis were: 1) it seems implausible that all samples used in Relax experiments contain impurities that are not present in any sample used for localization experiments, since both kinds of samples were fabricated in the same apparatus; and 2) we checked whether adding a term of the form $K(\varepsilon) \propto \varepsilon^{-2}$ to the interaction kernel could significantly decrease the value of $\kappa_{ee}$ obtained from fitting the data to the solution of \textsc{Eq.}~(\ref{Boltzmann}).  But those two arguments do not rule out another possibility, namely that both kinds of samples contain magnetic impurities with integer spin and with a magnetic anisotropy of the form $\mathcal{K}S_z^2$ in the impurity Hamiltonian \cite{privcomm}. Such a term is predicted in the presence of spin-orbit scattering, for magnetic impurities located close to the sample surface \cite{Zawadowski}. If the characteristic energy $\mathcal{K}$ satisfies $k_BT \ll \mathcal{K} < eU$, then such impurities would contribute to energy exchange but not to dephasing.  The contribution to $K(\varepsilon)$ from such impurities depends on both $\mathcal{K}$ and $B$, but is not expected to be of the form $K(\varepsilon) \propto \varepsilon^{-2}$.  In principle, the presence of such impurities should be detectable in experiments in the presence of a magnetic field. Indeed once $g \mu B \gg eU$, their contribution vanishes. The absence of visible magnetic field dependence in sample AgX20 seems to rule out this possibility.

In conclusion, the energy dependence of Coulomb interaction in disordered wires is well explained by theory. The intensity of the interaction, as deduced from phase coherence time measurements, is quantitatively in agreement with theory, whereas for energy relaxation, an unexplained discrepancy remains. A new version of the Relax experiment has demonstrated that this discrepancy is not due to the out-of-equilibrium situation. 

We gratefully acknowledge the contributions to this work by A.
Gougam, and helpful discussions with I.
Aleiner, G. Goeppert, H. Grabert and G. Montambaux.  Work at Saclay was supported in part by the EU Network DIENOW. Work at Michigan State University
was supported in part by the National Science Foundation under
grant DMR-0104178, and by the Keck Microfabrication Facility
supported by NSF DMR-9809688.


\begin{thebibliography}{00}
\bibitem{Lso} The widths $w$ of Ag wires determined from the best fits of magnetoresistance data to weak localization theory are often 10-15\% smaller than the widths determined
from electron microscope pictures.  This is probably due to the granularity of the Ag wires.
\bibitem{factor2} This expression for $\kappa_{ee}$ is half as large as the one used in \textsc{Refs.} \cite{PRB,PRLAnne,These Anne,TheseFred,PierreJLTP}.
\bibitem{kappaph facteur 2} In previous publications, we used $\kappa_{\textrm{ph}}=8~\textrm{ns}^{-1}\textrm{meV}^{-3}$ due to a factor of 2 error in extracting $B$ in \textsc{Eq.}~(\ref{tauphi}) from the Boltzmann equation (\textsc{Eq.}~(\ref{Boltzmann})).
\bibitem{ReservoirNote} Surprisingly, the values of the reservoir heating parameters that produce the best fits to the data are somewhat smaller than the values calculated from the reservoir heating model discussed in \cite{HennySchonenberger}.
\bibitem{R not measured} The resistivity of the wire could not be accessed in the experiment. It has been estimated by comparison with AgXV40, which was fabricated simultaneously, assuming equal diffusion constants.
\end{thebibliography}
\end{document}